\documentclass[dvips]{article}
\usepackage{supertabular,lscape,epsfig}
\usepackage{graphicx}
\usepackage{amssymb}
\usepackage{amsmath}
\textheight=22cm \DeclareSymbolFont{ppa}{OT1}{ppl}{m}{it}
\DeclareMathSymbol{\vv}{\mathalpha}{ppa}{'166}

minus 2.3mu \thickmuskip = 2.6mu plus 2mu minus 2.6mu

\begin{document}

\newcommand{\dd}{\,{\rm d}}
\newcommand{\ie}{{\it i.e.},\,}
\newcommand{\etal}{{\it et al.\ }}
\newcommand{\eg}{{\it e.g.},\,}
\newcommand{\cf}{{\it cf.\ }}
\newcommand{\vs}{{\it vs.\ }}
\newcommand{\zdot}{\makebox[0pt][l]{.}}
\newcommand{\up}[1]{\ifmmode^{\rm #1}\else$^{\rm #1}$\fi}
\newcommand{\dn}[1]{\ifmmode_{\rm #1}\else$_{\rm #1}$\fi}
\newcommand{\upd}{\up{d}}
\newcommand{\uph}{\up{h}}
\newcommand{\upm}{\up{m}}
\newcommand{\ups}{\up{s}}
\newcommand{\arcd}{\ifmmode^{\circ}\else$^{\circ}$\fi}
\newcommand{\arcm}{\ifmmode{'}\else$'$\fi}
\newcommand{\arcs}{\ifmmode{''}\else$''$\fi}
\newcommand{\MS}{{\rm M}\ifmmode_{\odot}\else$_{\odot}$\fi}
\newcommand{\RS}{{\rm R}\ifmmode_{\odot}\else$_{\odot}$\fi}
\newcommand{\LS}{{\rm L}\ifmmode_{\odot}\else$_{\odot}$\fi}

\newcommand{\Abstract}[2]{{\footnotesize\begin{center}ABSTRACT\end{center}
\vspace{1mm}\par#1\par \noindent {~}{\it #2}}}

\newcommand{\TabCap}[2]{\begin{center}\parbox[t]{#1}{\begin{center}
  \small {\spaceskip 2pt plus 1pt minus 1pt T a b l e}
  \refstepcounter{table}\thetable \\[2mm]
  \footnotesize #2 \end{center}}\end{center}}

\newcommand{\TableSep}[2]{\begin{table}[p]\vspace{#1}
\TabCap{#2}\end{table}}

\newcommand{\FigCap}[1]{\footnotesize\par\noindent Fig.\  %
  \refstepcounter{figure}\thefigure. #1\par}

\newcommand{\TableFont}{\footnotesize}
\newcommand{\TableFontIt}{\ttit}
\newcommand{\SetTableFont}[1]{\renewcommand{\TableFont}{#1}}

\newcommand{\MakeTable}[4]{\begin{table}[htb]\TabCap{#2}{#3}
  \begin{center} \TableFont \begin{tabular}{#1} #4
  \end{tabular}\end{center}\end{table}}

\newcommand{\MakeTableSep}[4]{\begin{table}[p]\TabCap{#2}{#3}
  \begin{center} \TableFont \begin{tabular}{#1} #4
  \end{tabular}\end{center}\end{table}}

\newenvironment{references}%
{ \footnotesize \frenchspacing
\renewcommand{\thesection}{}
\renewcommand{\in}{{\rm in }}
\renewcommand{\AA}{Astron.\ Astrophys.}
\newcommand{\AAS}{Astron.~Astrophys.~Suppl.~Ser.}
\newcommand{\ApJ}{Astrophys.\ J.}
\newcommand{\ApJS}{Astrophys.\ J.~Suppl.~Ser.}
\newcommand{\ApJL}{Astrophys.\ J.~Letters}
\newcommand{\AJ}{Astron.\ J.}
\newcommand{\IBVS}{IBVS}
\newcommand{\PASP}{P.A.S.P.}
\newcommand{\Acta}{Acta Astron.}
\newcommand{\MNRAS}{MNRAS}
\renewcommand{\and}{{\rm and }}
\section{{\rm REFERENCES}}
\sloppy \hyphenpenalty10000
\begin{list}{}{\leftmargin1cm\listparindent-1cm
\itemindent\listparindent\parsep0pt\itemsep0pt}}%
{\end{list}\vspace{2mm}}

\def\TYLDA{~}
\newlength{\DW}
\settowidth{\DW}{0}
\newcommand{\dw}{\hspace{\DW}}

\newcommand{\refitem}[5]{\item[]{#1} #2%
\def\REFARG{#3}\ifx\REFARG\TYLDA\else, {\it#3}\fi
\def\REFARG{#4}\ifx\REFARG\TYLDA\else, {\bf#4}\fi
\def\REFARG{#5}\ifx\REFARG\TYLDA\else, {#5}\fi.}

\newcommand{\Section}[1]{\section{#1}}
\newcommand{\Subsection}[1]{\subsection{#1}}
\newcommand{\Acknow}[1]{\par\vspace{5mm}{\bf Acknowledgements.} #1}
\pagestyle{myheadings}

\newfont{\bb}{ptmbi8t at 12pt}
\newcommand{\xrule}{\rule{0pt}{2.5ex}}
\newcommand{\xxrule}{\rule[-1.8ex]{0pt}{4.5ex}}
\def\thefootnote{\fnsymbol{footnote}}

\begin{center}
 {\Large\bf
  Variable Stars in the Field of the Open Cluster NGC 2204}
  \vskip1cm
  {\bf M.~~R~o~z~y~c~z~k~a$^1$,
      ~~J.~~K~a~l~u~z~n~y$^1$,
      ~~W.~~K~r~z~e~m~i~n~s~k~i$^2$,
      ~~and~~B.~~M~a~z~u~r$^{1}$\\}
  \vskip3mm {
  $^1$Nicolaus Copernicus Astronomical Center,
     ul. Bartycka 18, 00-716 Warsaw, Poland\\
     e-mail: (mnr,jka,batka@camk.edu.pl)\\
  $^2$Las Campanas Observatory, Casilla 601, La Serena, Chile,\\
     e-mail: (wojtek@lco.cl)}\\
\end{center}

\Abstract{ We present the results of a variable stars search in
the field of the old open cluster NGC 2204. Five new variables were found, 
four of them being eclipsing binaries. The sample includes a detached binary 
located at the turnoff, a W UMa -- type system, and an interesting
detached low-mass binary with a period of $0.45\upd$ which, however, 
is a foreground object.
We provide $V$-light curves and finder charts for all variables together with 
color-magnitude diagrams of the cluster. For four variables incomplete 
$I$-light curves are also provided.}
{\bf Key words:} {\it open clusters and associations: individual:
NGC 2204 - binaries: eclipsing}
\section{Introduction} \label{sect:intro}
NGC 2204 (Mel 44) is a rich but rather loose cluster in Canis Major, 
centered at 
$(l,b)=(226.01,-16.11)$. It is located $4$ kpc 
away from the Sun, 11.1 kpc away from the galactic 
center and 1.1 kpc below the galactic plane (Kassis et al. 1997). 
At that heliocentric distance 
its angular diameter of 15\arcm\ (Hawarden 1976) corresponds to 
the linear diameter of 
$\sim17$ pc. It is only weakly reddened, with estimates of 
$E(B-V)$ ranging from 0.07  (Mermilliod and Mayor 2007) to 0.13 
(Kassis et al. 1997). Maps of interstellar reddening by Schlegel et al. (1998)
imply that reddening in the cluster should not exceed $E(B-V)=0.10$.
Estimates of its age vary from 0.78 Gyr (Piatti \etal 2003) to 
2.5 Gyr (von Hippel 2005). Various measurements 
of [Fe/H] have yielded values between $-0.47$ (Hou \etal 2002) 
and $-0.32$ (Krusberg  and Chaboyer 2006), consistently showing that 
the cluster is metal-deficient.

Deep CCD photometry of NGC 2204 was obtained by Kassis \etal (1997), who also performed
isochrone fitting. They derived an age of $1.6^{+0.9}_{-0.3}$ Gyr, with uncertainties 
ascribed mainly to the poorly known metallicity. Mermilliod and Mayor (2007) obtained 
spectra for 25 red giants belonging to the cluster. The sample included all red clump 
stars together with those brighter than the clump. Based on these data, they found the 
mean radial velocity of the cluster to be equal to $91.38\pm0.30$ km s$^{-1}$. 

No survey of variable stars in NGC 2204 has been reported so far. The present study 
is based on archival observations performed within the long-term program of CCD search 
for short-period variables in selected star clusters (see e.g.
Mazur \etal 1999 and references therein). Particularly interesting and valuable among 
those variables are detached eclipsing binaries, which at the moment enable the most 
accurate determinations of stellar masses, radii and luminosities, thus providing an 
excellent check on stellar evolution theory and a means for direct distance determination.
\begin{figure}[t]
 \begin{center}
  \includegraphics[width=\textwidth, bb = 46 190 564 694]{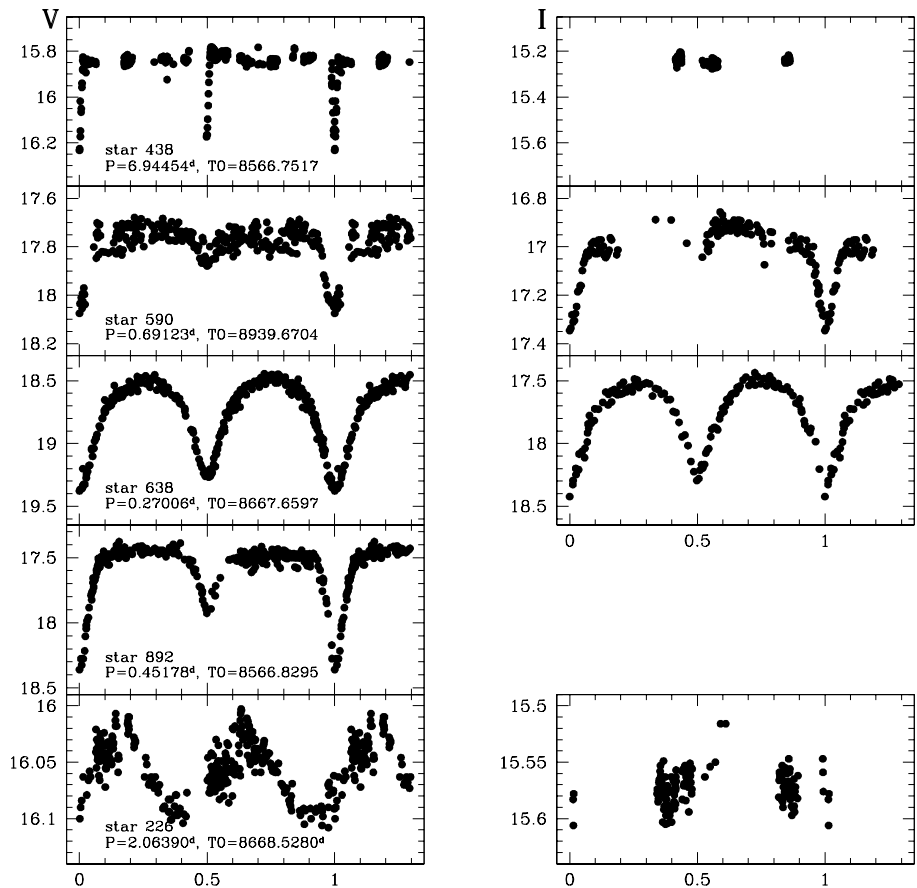}
  \caption {Phased light curves. T0 are moments of minima on JD-2440000.
   The star V892 was visible in only one $I$-frame
   (the single one taken with Tek \#3).}
  \label{fig:lcurves}
 \end{center}
\end{figure}
\section{Observations and Data Reduction}\label{sect:observations}
The data were collected with the 1-m Swope telescope at Las Campanas 
Observatory on 24 nights between October 1991 and January 1993. The 
cluster was observed with three different CCD cameras: 
Tektronix \#1 ($1024\times1024$ pixels; 12 arcmin FOV; 0.70 arcsec/pixel), 
Tektronix \#2 ($1024\times1024$ pixels; 10.4 arcmin FOV; 0.61 arcsec/pixel) and 
Tektronix \#3 ($2048\times2048$ pixels; 21 arcmin FOV; 0.61 arcsec/pixel). 
Altogether, 485 frames were taken (310, 136, 29 and 10 in $V$, $I$, $B$ and 
$U$ filters, respectively). The exposure times ranged from 10 to 900 s.

The preliminary processing of the raw data was performed under 
IRAF\footnote{IRAF is distributed by the National Optical Astronomy
Observatories, which are operated by the AURA, Inc., under cooperative 
agreement with the NSF.}. 
Tektronix \#3 frames were cropped to $1200\times1200$ pixels (12 arcmin FOV). 
All frames were bias-subtracted and flattened with median-averaged
sky flats. Because of problems with the filter the flattening did not 
work well for $U$-frames, and the resulting $U$-magnitudes
are less reliable than $B$, $V$ or $I$ ones.  

The photometry was performed with the \textsc{Daophot/Allstar} package 
(Stetson 1987). The reduction procedure started from the 
identification of stars with subroutine \textsc{Find}, followed by aperture 
photometry with subroutine \textsc{Phot}. Next, based on about a hundred isolated 
stars, a PSF varying quadratically with $(x,y)$ coordinates was constructed for
each frame,
and used for the profile photometry with subroutine \textsc{Allstar}. The images 
were inspected visually, and for each camera and each filter one of the best frames 
was chosen as a template. Instrumental magnitudes of template stars were transformed 
to the standard system using the data of Kassis \etal (1997) available in the WEBDA 
database. 
For each frame taken with the same camera and filter the template stars were 
identified by means of transformations of $(x,y)$ coordinates, 
and \textsc{Allstar} photometry for those frames was transformed to the standard 
system with the template serving as an intermediary. Finally, $\alpha_\mathrm{2000}$ 
and $\delta_\mathrm{2000}$ were determined for all stars found in the $V$-filter. The 
transformation from rectangular to equatorial coordinates was based on 1557 stars from 
the USNO-A2 catalogue with  
$06^\mathrm{h} 16^{\mathrm{m}\,} 01^{\mathrm{s}} > \alpha_\mathrm{2000} >
 06^\mathrm{h} 14^{\mathrm{m}\,} 57^{\mathrm{s}}$, and  
$-18^{\circ} 31\arcm 43\arcs > \delta_\mathrm{2000} > 
 -18^{\circ} 46\arcm 46\arcs$. 
\section{Variable Stars}\label{sect:vars}
\begin{figure}[h]
 \includegraphics[width=60 mm, bb = 45 180 570 695]{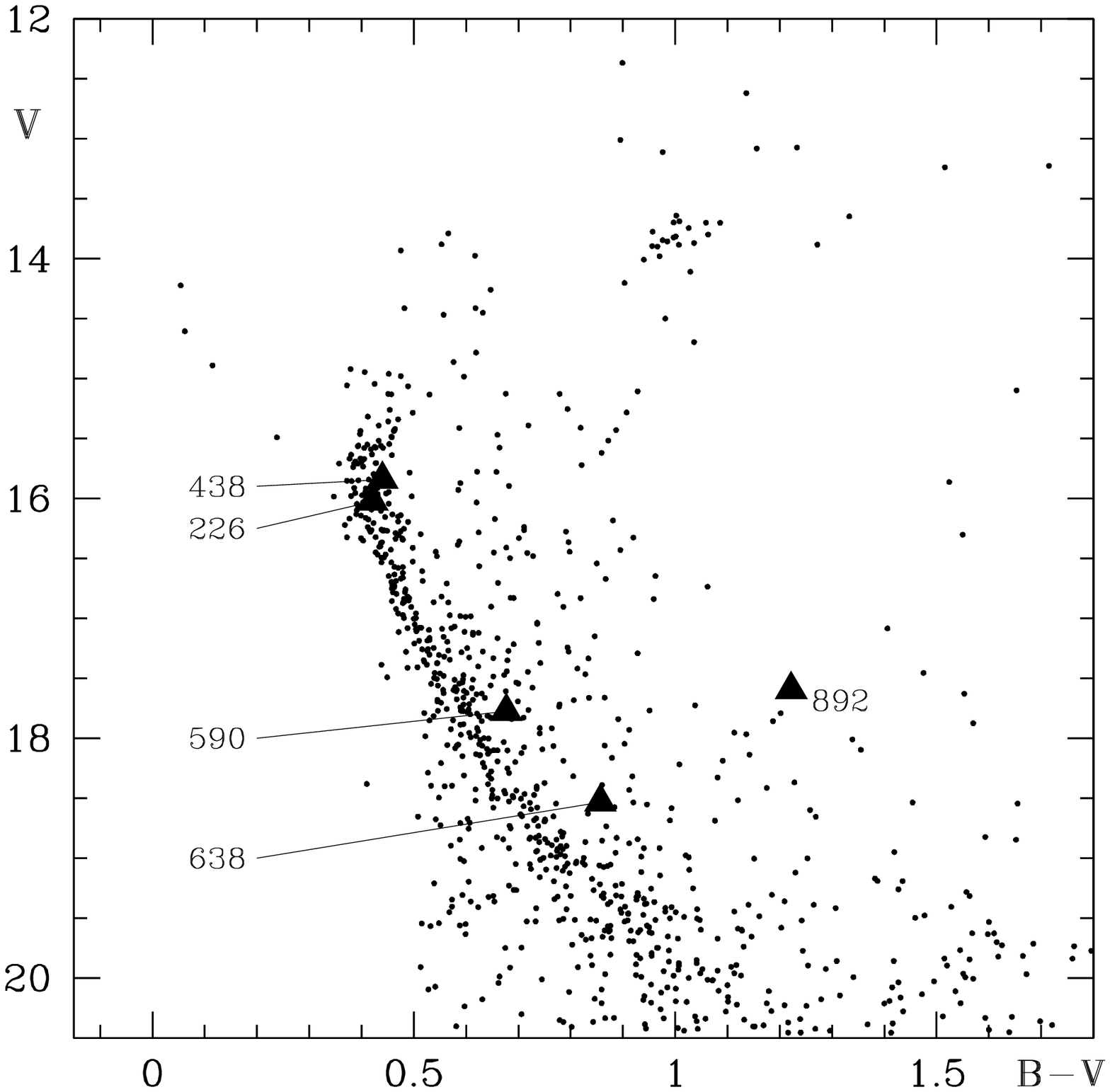}
 \includegraphics[width=60 mm, bb = 45 180 570 695]{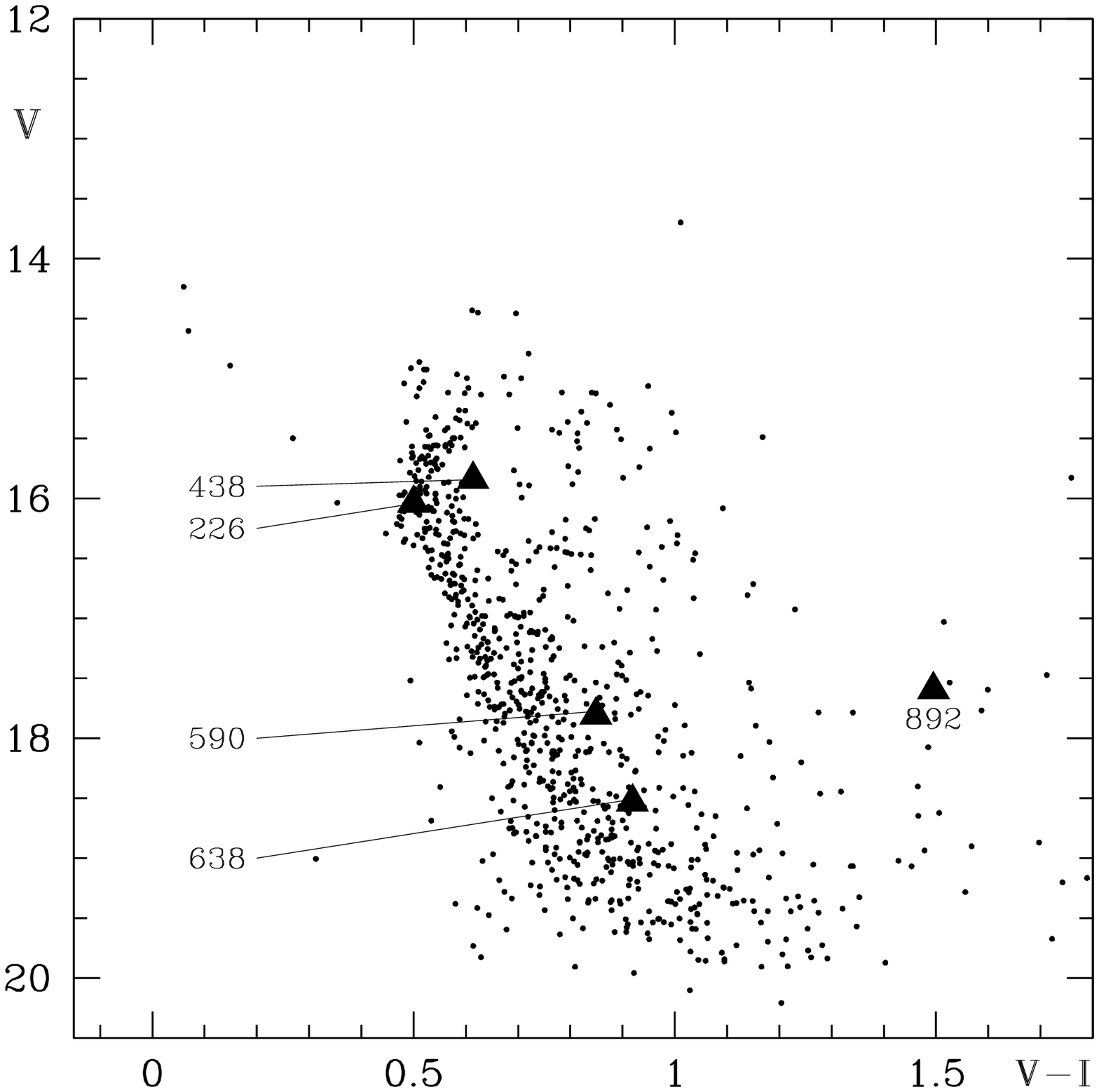}
 \caption {Color-magnitude diagrams of NGC 2204 with locations
  of the  discovered variables. Most stars brighter than ~14 mag in $I$ were
  overexposed.}
 \label{fig:cmds}
\end{figure}

\begin{figure}
 \begin{center}
  \includegraphics*[width=39.5 mm, bb = 194 287 416 507]{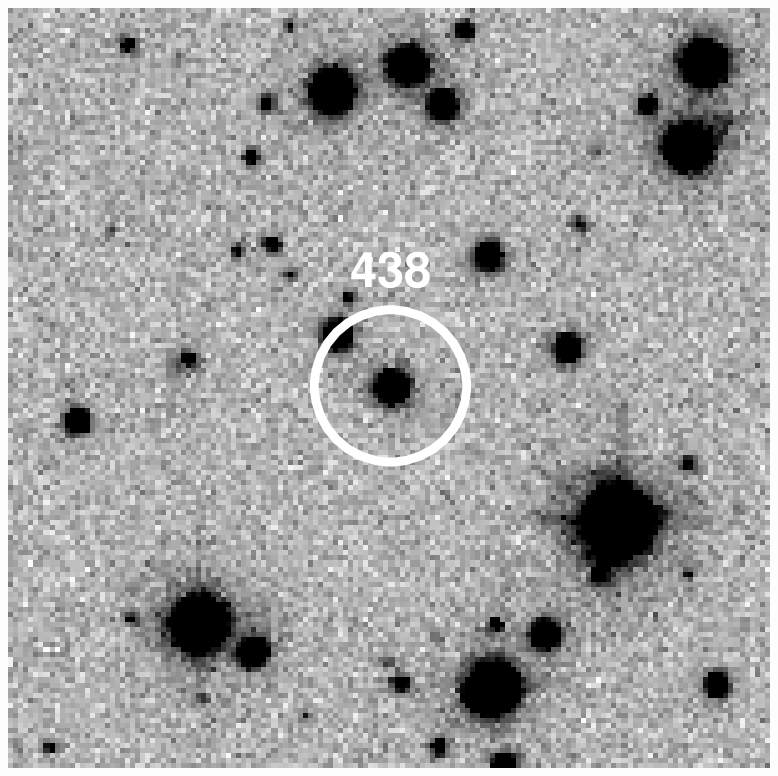}
  \includegraphics*[width=39.5 mm, bb = 194 287 416 507]{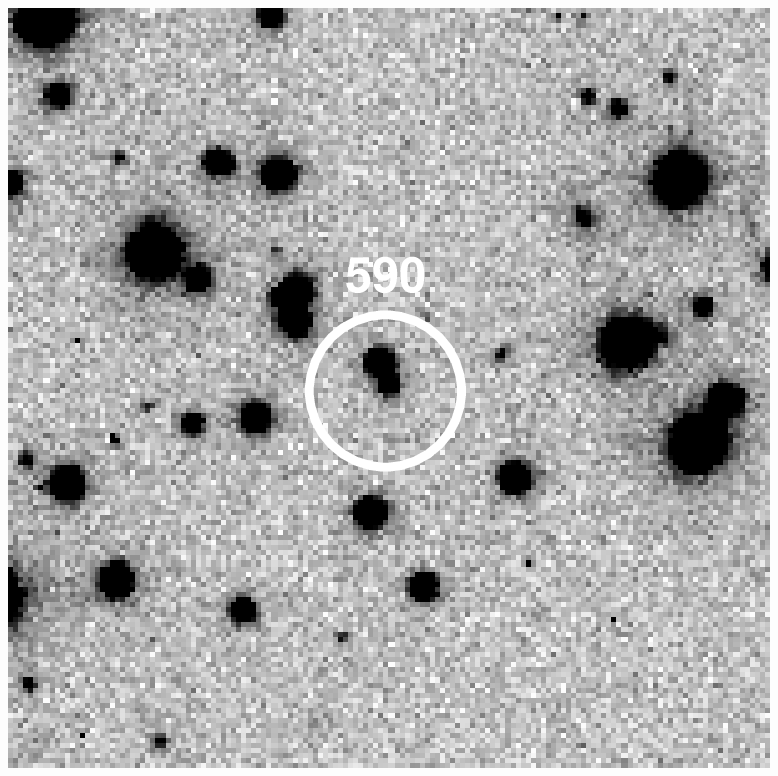}
  \includegraphics*[width=39.5 mm, bb = 194 287 416 507]{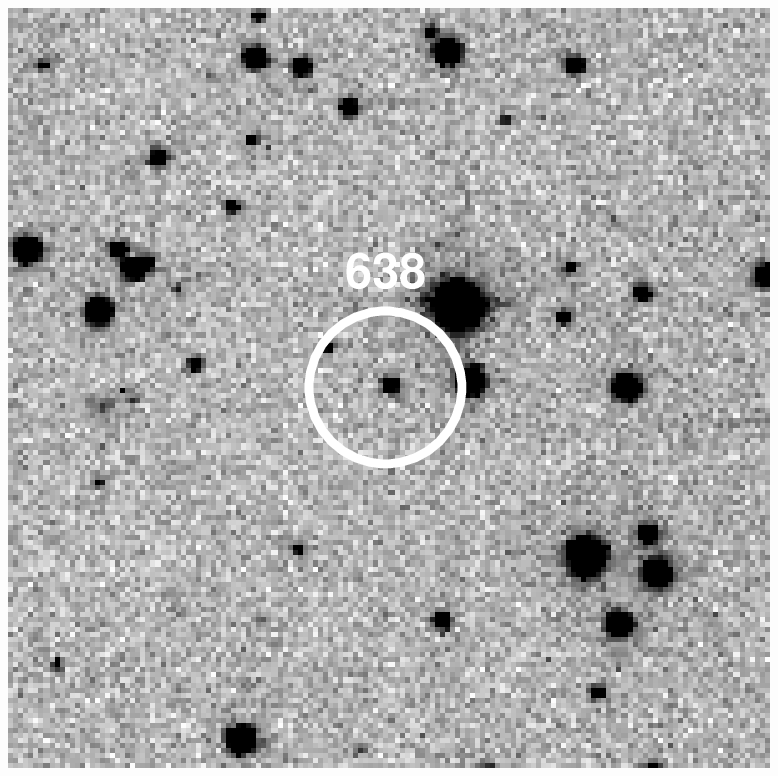}
  \includegraphics*[width=39.5 mm, bb = 194 287 416 507]{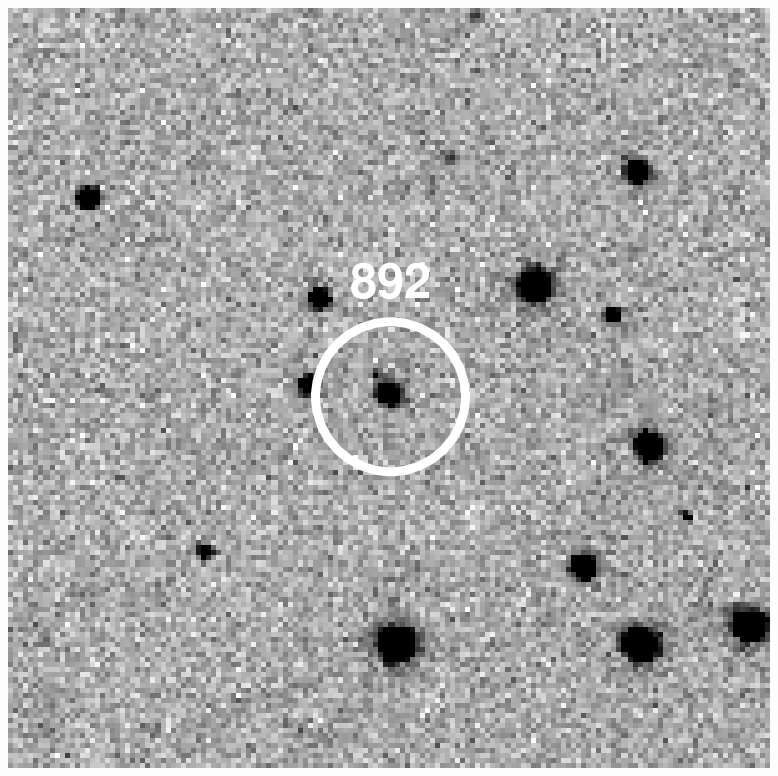}
  \includegraphics*[width=39.5 mm, bb = 196 306 415 527]{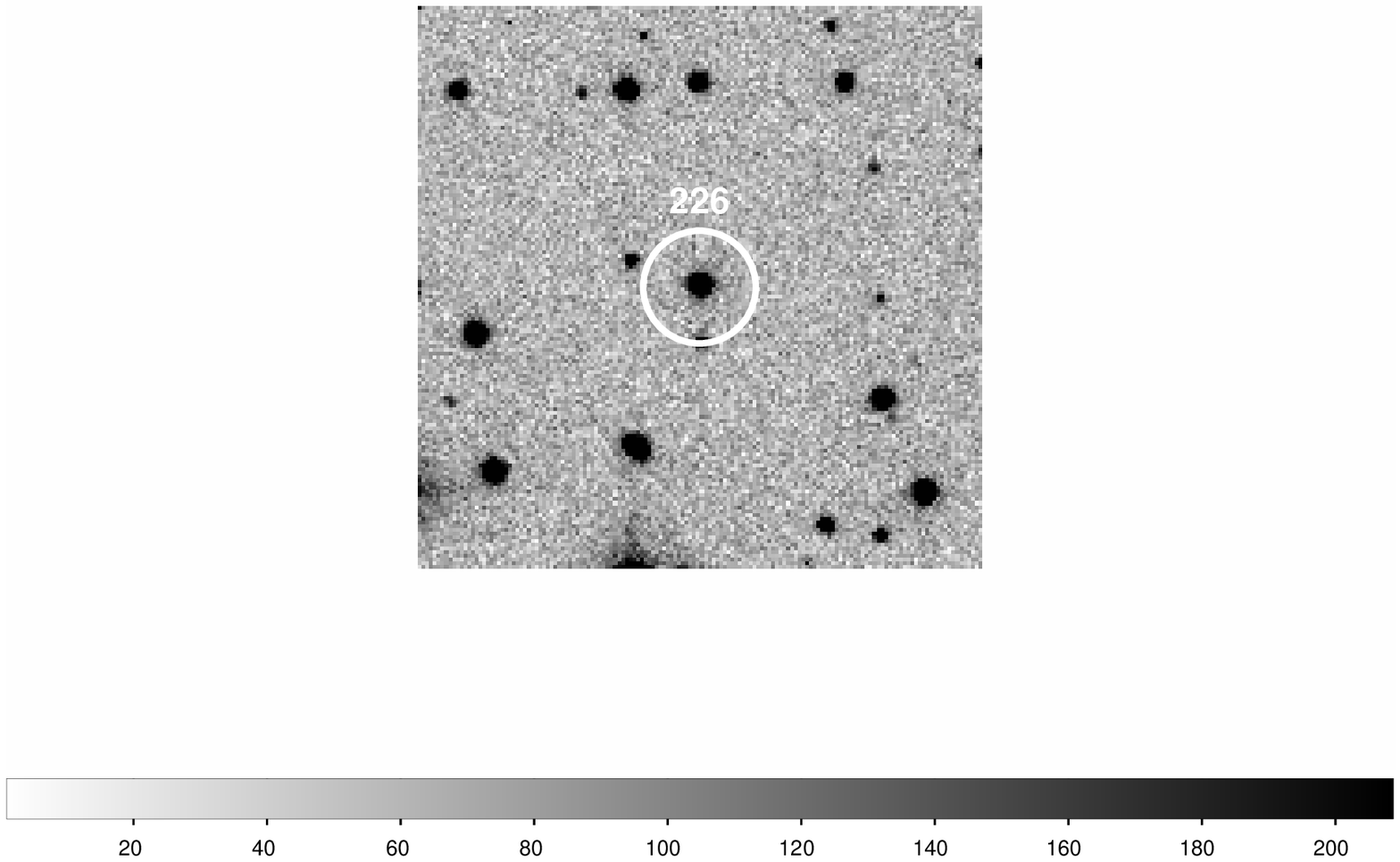}
  \includegraphics*[width=39.5 mm, bb = 196 306 415 526]{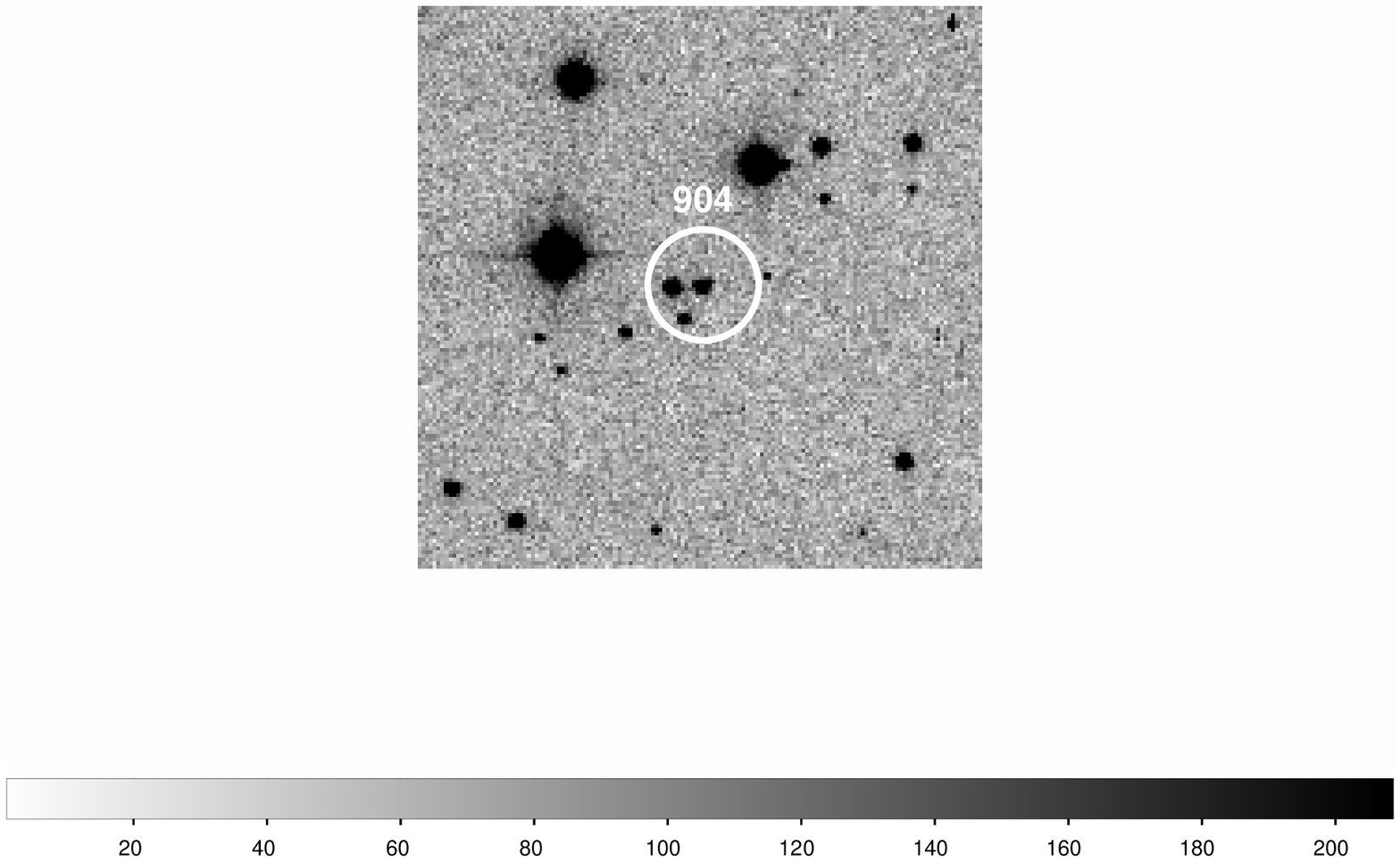}
   \caption {The $V$-band finder charts for the new variables.
    Each chart is $90\arcs$ on a side with North up and East to the left.}
   \label{fig:charts}
 \end{center}
 \vskip -3 mm
\end{figure}

In order to effectively use the available information, a field of 
$12\arcm\times12\arcm$ 
was analyzed, so that only the photometry from Tektronix \#2 was missing for some stars. 
For that field all photometric data were combined into a database, and searched for
periodicities with the program TATRY kindly provided by Alex Schwarzenberg-Czerny. 
This program is suitable for the detection of various types of variables based on 
the analysis of variance method (Schwarzenberg-Czerny 1996). Four evident variables
were found, whose light curves are shown in Fig. \ref{fig:lcurves}. An inspection 
of cluster's color-magnitude diagrams (Fig. \ref{fig:cmds}) suggests that stars 
V438, V590 and V638 indeed belong to NGC 2204. They are located on or 
slightly above the cluster main-sequence, which is consistent 
with their binary nature. The star V892 must be 
a foreground object as it is located far redwards of the cluster main sequence. 
Finder charts for the detected variables are shown in 
Fig. \ref{fig:charts}.

Star V438 is a clearly detached MS binary on a circular orbit. Within observational 
errors its $V$- curve is flat between the minima. The minima are deep and similar
in shape, indicating that the system is seen nearly edge-on, and 
its components are almost identical. 
Unfortunately, none of the minima occurred  during relatively scarce $I$-observations. 
In Fig. \ref{fig:cmds} star V438 is located practically at the turnoff.
This object seems to 
be well suited for spectroscopic observations which would allow for an accurate 
determination of the masses of its components and a much better estimate of the
age and distance of NGC 2204. 
\begin{table}
\begin{center}
\caption{\small Coordinates and photometric parameters of variables from NGC~2204 
field. $(B-V)$ and $(V-I)$ are given at maximum $V$.}
{\small
\begin{tabular}{cccccccc}
\hline
Star  &RA$_{\rm 2000}$ &Dec$_{\rm 2000}$& $V_{\rm MAX}$& $\Delta V$& $(B-V)$
      & $(V-I)$&Type\\
\hline
438       &06:15:33.85& -18:39:20.2&  15.84& 0.45&    0.55& 0.61&Ecl\\
590       &06:15:29.92& -18:40:40.1&  17.75& 0.38&    0.79& 0.85&Ecl\\
638       &06:15:23.40& -18:41:14.2&  18.45& 0.93&    0.94& 0.92&Ecl-EW\\
892       &06:15:55.42& -18:44:51.7&  17.45& 0.95&    1.17& 1.50&Ecl\\
226       &06:15:46.10& -18:37:6.70&  16.04& 0.08&    0.52& 0.50&Ell? \\
904       &06:15:17.82& -18:45:07.1&  18.28& $<0.05$&  0.28$^\star$& 0.91&?  \\
\hline
\end{tabular}}
\end{center} 
\vskip -2 mm
{\footnotesize$^\star(U-B)=-0.9$}
\end{table}

Star V590 is blended with a brighter but resolved neighbor, resulting 
in a rather poor photometry. 
Judging from the $V$-curve it is another detached binary, however inclined to the 
observer at an angle significantly different from 90\arcd. The $I$-curve does not 
entirely cover the secondary minimum, but it seems to indicate effects of proximity, 
consistent with the short period of this system. 

Star V638 is a contact binary with primary minima $\sim1$ mag deep, and secondary 
minima only $\sim0.1$ mag shallower. Large amplitude of light variations 
indicates mass ratio close to unity  and inclination close to 90 deg. 
Using empirical calibration of Rucinski and Duerbeck (1997) 
we have estimated absolute magnitude of the variable at $M_{\rm V}=5.27$.
With $V_{max}=18.53$ this implies an apparent distance modulus 
$(m-M)_{\rm V}=13.26$ what is very close to distance modulus of the cluster 
estimated at $(m-M)_{\rm V}=13.28$ by Mermilliod \& Mayor (2007).
Hence, star V638 is a very likely member of NGC~2204.    

Star V892 -- a detached binary 
with $B-V=1.17$ and a period of $0.45\upd$ -- is the most interesting among the new 
variables. Its short period together with the undetectable
proximity effects suggest that it is composed of MS stars. Such system can be located 
high above the MS of the cluster only if it is a foreground object. As the primary 
minimum is very deep, it must be seen nearly edge-on. A much shallower secondary 
minimum indicates a significant difference in surface brightness of the components. 
To roughly estimate the parameters of the primary we can neglect the contribution of the 
secondary and assume that the reddening of the system is two times smaller than that 
of NGC 2204. We get the true $B-V$ of $\sim 1.25$, {\it i.e.} an effective temperature 
of $\sim$4300 K (spectral type K6-K7), which on the Main Sequence translates into 
${\cal M}_\star\approx0.6\MS$ and $M_V\approx8$ (Sekiguchi and Fukugita 1999; 
Baraffe \etal 1998). Since $V$ is equal to $17.5$, star V892 should be located 
at a distance of about 1 kpc from the Sun, consistent with the assumed value of
the reddening. An equally rough estimate of the parameters of the secondary yields
$M_V\approx8.5$ and spectral type K7-K8. 

Star V226 shows a sinusoidal light curve with a full amplitude of about 0.08 mag 
in the $V$ band. It may be an ellipsoidal variable, but our data
are not sufficient for an unambiguous classification of this object.

Finally, we found a potentially interesting blue object - star V904 
with $(U-B) = -0.9$, $(B-V)= 0.28$ and $V=18.3$ which may be 
variable with a  period of $\sim1.4\upd$
or $\sim2.8\upd$ and an amplitude smaller than $\sim0.05$ mag. 
We could not identify
it with any cataloged X-ray source, and in the future it might be 
worthwhile to take 
its spectrum. One may note that the strongly blue $U-B$ color of star V904 is
incompatible with its $B-V$ and especially $V-I$ color. This may
indicate that we are dealing with a composite object. Finder charts for stars V226
and V904 are also shown in Fig. \ref{fig:charts}.

None of the four blue straggler candidates visible in Fig. 2 showed any evidence
for variability with a full amplitude exceeding 0.05 mag.  

\section{Discussion}\label{sect:discussion}
We performed a search for variable stars in the field of the old open cluster 
NGC 2204 which yielded four new eclipsing binaries. 
One of them (star V438) is located at the turnoff, and may be used for an accurate 
determination of the age and distance of the cluster. Another one (star V892, a 
foreground object) is an interesting detached binary, most likely composed of two 
late K-type MS stars. Such systems are very rare -- Ribas (2006) lists only 14 members 
of double lined binaries with masses below $0.8\MS$ -- and very puzzling, as their 
observed radii disagree with theoretical predictions. In the $0.4-0.8\MS$ range the  
models predict values which are consistently too small by 5 -– 15\%. Since the current 
accuracy of the measurements reaches a few percent in both mass and radius, these 
differences are undoubtedly significant. Thus, star V892 is potentially very valuable. 
At $V=17.5$ mag its spectrum could be easily obtained on a modern large 
telescope. If
our estimates are correct, and lines of both components are visible, it would help 
to solve the problem of discrepancies on the lower Main Sequence.   
\section{Acknowledgments}
This research has made use of the WEBDA database, operated at the Institute 
for Astronomy of the University of Vienna. MR thanks Pawe\l\ Pietrukowicz and Wojtek 
Pych for numerous advices concerning the data reduction process. JK and BM were 
supported by the grant 1P03D 001 28 from the Polish Ministry of Science. Research 
of JK is also supported by the Foundation for Polish Science through the grant MISTRZ.

\begin{references}
%
\refitem{Baraffe, I., Chabrier, G., Allard, F., and Hauschildt, P.~H.}
        {1998}{\AA}{337}{403}
\refitem{Hawarden, T.~G.}{1976}{\MNRAS}{174}{225}
\refitem{Hou, J.-L., Chang, R.-X., and Chen, L.}
        {2002}{Chinese J. Astron. Astrophys.}{2}{17}
\refitem{Kassis, M., Janes, K.~A., Friel, E.~D., and Phelps, R.~L.}
        {1997}{\AJ}{113}{1723}
\refitem{Krusberg, Z.~A.~C., and Chaboyer, B.}{2006}{\AJ}{131}{1565}
\refitem{Mazur, B., Krzemi\'nski, W., and Kaluzny, J.}{1999}{\Acta}{49}{551}
\refitem{Mermilliod, J.-C., and Mayor, M.}{2007}{\AA}{470}{919}
\refitem{Piatti, A.~E., Claria, J.~J., and Ahumada, A.~V.}{2003}{\MNRAS}{346}{390}
\refitem{Ribas, I.}{2006}{Astrophys. Sp. Sci}{304}{89}
\refitem{Rucinski, S.~M., and Duerbeck, H.~W.}{1997}{\PASP}{109}{1340}
\refitem{Schlegel D.J., Finkbeiner, D.P., Davis, M.}{1998}{\ApJ}{500}{525}
\refitem{Schwarzenberg-Czerny, A.}{1996}{\ApJL}{460}{L107}
\refitem{Sekiguchi, M., and Fukugita, M.}{1999}{\AJ}{120}{1072}
\refitem{Stetson, P.~B.}{1987}{\PASP}{99}{191}
\refitem{von Hippel, T.}{2005}{\ApJ}{622}{565}
%
\end{references}
\end{document}